\documentstyle[12pt]{article}
\newcommand{\D}{\displaystyle}
\newcommand{\uu}{\widehat{\stackrel{\rightarrow}{u}}}
\newcommand{\qq}{\widehat{\stackrel{\rightarrow}{Q}}}
\newcommand{\vv}{{\stackrel{\rightarrow}{v}}}

\newcommand{\h}{\hbar}

\begin{document}
\begin{center}
{\bf An Alternative Commutation Relation Between Position And Momentum Operators of Massless Particles}\\
\vspace{0.5cm}

{\bf H. Razmi}

\vspace{0.5cm}

{\it \small Department of Physics, School of sciences \\
Tarbiat Modarres University, P.O.Box 14155-4838\\
Tehran, I.R. Iran }\\
{\bf Abstract}
\end{center}
It seems that the problem of finding a suitable position operator for photon
has been solved in a recently published work which is based on a new commutation
relation between position and momentum operators of massless particles[1].
Although the authors of [1] have presented a classical reasoning in favour of
the new commutator,here we are going to find this new commutator based on a
quantum mechanical argument.\\

\newpage
\subsection*{Introduction:}
The relation $[\hat{Q}_i,\hat{P}_j]=i\h \delta_{ij}\hat{1}$ is generally 
accepted as the canonical commutation relation between position and momentum 
operators of any particle. For massive particles, it can be derived by means 
of symmetry group transformations and the related Lie algebra commutators [2-4]. 
The derivation is based on the transformation property of the velocity 
operator $\uu=\frac{d\qq}{dt}$, which can be transformed to 
$\uu-\vv \hat{1}$ (where $\vv$ is an infinitesimal velocity parameter)
by operation of boost generators. However for massless particles e.g. photons, 
this transformation is not valid. The reason originates from the fact that the 
velocity of light is constant and cannot be altered by any transformation (it 
is a well-known result of Poincare transformations that there is no rest frame 
for massless particles). Consequently the derivation of the commutation 
relation $[\hat{Q}_i, \hat{P}_j]=i\h \delta_{ij}\hat{1}$ based on this 
velocity transformation cannot be valid for massless particles. 

\subsection*{The Commutation Relation Between Position And Momentum Operators 
For Massless Particles:}

Here we are going to deal with the commutation relation of the operators 
$\hat{Q}$ and $\hat{P}$ for massless particles. The Heisenberg equation of 
motion for position operator, based on the fact that the Hamiltonian operator 
is the generator of time translation, gives:
\begin{equation}
[\hat{Q_i}, \hat{H}]=i\h \frac{d\hat{Q}_i}{dt}
\end{equation}
and the momentum operator satisfies 
\begin{equation}
\hat{P}_i= \left(\frac{\hat{H}}{c^2} \right )
\left(\frac{d\hat{Q}_i}{dt} \right )
\end{equation}
The employment of (1) and (2) in the commutator of 
$\hat{Q}_i$ and $\hat{P}_j$ readily gives: 
\begin{equation}
\begin{array}{ll}
[\hat{Q}_i, \hat{P}_i] & =
\left[ \hat{Q}_i, \frac{\hat{H}}{c^2} \frac{d\hat{Q}_i}{dt} \right]=
\frac{1}{c^2} [\hat{Q}_i, \hat{H}] \frac{d\hat{Q}_j}{dt}+\frac{\hat{H}}{c^2} 
\left[ \hat{Q}_i, \frac{d\hat{Q}_j}{dt} \right] \\
& =\frac{i\h}{c^2} \left( \frac{d\hat{Q}_i}{dt}\right) 
\left( \frac{d\hat{Q}_j}{dt}\right) + \frac{\hat{H}}{c^2} 
\left[ \hat{Q}_i, \frac{d\hat{Q}_j}{dt} \right] 
\end{array}
\end{equation}
This relation is valid for both massless and massive particles. For massless 
particles the velocity is a constant operator and hence we have (see section 
A):
\begin{equation}
\left[ \hat{Q}_i, \frac{d\hat{Q}_j}{dt} \right] =0
\end{equation}
Therefore (3) and (4) give the following commutation relation for massless 
particles: 
\begin{equation}
[\hat{Q}_i,\hat{P}_j]= \frac{i\h}{c^2} 
\left( \frac{d\hat{Q}_i}{dt}\right) 
\left( \frac{d\hat{Q}_j}{dt}\right) 
\end{equation}
By means of (2) the above result can be also written in the following form:
\begin{equation}
[\hat{Q}_i,\hat{P}_j]=i\h c^2 H^{-2} \hat{P}_i \hat{P}_j
\end{equation} 
\subsection*{Outlook:}
In addition to the solution of the problem of finding a suitable position
operator for photon [1],the canonical quantum field theory of radiation
based on this new commutation relation seems to lead to interesting results[5].
\newpage
\noindent
{\bf Section A:} \\
The commutators of a Lie Algebra is (must be) independent of how we choose the 
coordinate system and one can always choose a coordinate system in which 
j-direction identifies with the direction of photon velocity vector operator 
and therefore $\frac{d\hat{Q}_j}{dt}$ in our relation (4) is identified with 
$c\hat{1}$ and the relation (4) is trivially proved. \\
Also, we can algebraically prove the relation (4): \\
$\left \{ \begin{array}{lll}
\D{\sum_{j=1}^3} \left(\frac{dQ_j}{dt} \right)^2 =c^2 \hat{1} & 
\hspace*{2cm} & \\
& & (massless particles)\\ 
\D{\frac{d^2 Q_j}{dt^2}=0} & & 
\end{array}\right .$\\
$$\Rightarrow \sum_{j=1}^3 \left [ \hat{Q}_i, 
\left ( \frac{d\hat{Q}_j}{dt} \right)^2 \right]=0 
\Rightarrow \sum_{j=1}^3 \left [ \hat{Q}_i, 
 \frac{d\hat{Q}_j}{dt} \right]\frac{d\hat{Q}_j}{dt}+\frac{d\hat{Q}_j}{dt} 
\left[ \hat{Q}_i, \frac{d\hat{Q}_j}{dt} \right]=0 \ \ \ \ (I)$$
But: 
$$\frac{d}{dt} \left [ \hat{Q}_i, \frac{d\hat{Q}_j}{dt} \right] =
\left [ \frac{d\hat{Q}_i}{dt}, \frac{d\hat{Q}_j}{dt} \right]+ 
 \left [ \hat{Q}_i, \frac{d^2\hat{Q}_j}{dt^2} \right] = [\hat{H}^{-1} 
\hat{P}_i, \hat{H}^{-1} \hat{P}_j ]=0$$
Therefore $\left[ \hat{Q}_i, \frac{d\hat{Q}_j}{dt} \right]$ is a constant 
operator or at most a function of the operators $\hat{H}$ and $\hat{P}$
 (note that 
we are speaking about free particles). \\
Since $\frac{d\hat{Q}_j}{dt}=\hat{H}^{-1} \hat{P}_j$ and according to the 
above result: \\
$\left[ [\hat{Q}_i, \frac{d\hat{Q}_j}{dt} ] , \frac{d\hat{Q}_j}{dt} \right] =0 
\hspace*{10cm} (II)$\\
Using (I) and (II):\\
$\D{\sum_{j=1}^3} \left ( \left [ \hat{Q}_i, \frac{d\hat{Q}_j}{dt} \right] 
\frac{d\hat{Q}_j}{dt} \right) =0$\\
But, generally, this is impossible unless: \\
$\left[ \hat{Q}_i, \frac{d\hat{Q}_j}{dt} \right]=0$

\newpage 

\begin{center}
{\bf REFERENCES} 
\end{center}

\begin{itemize}
\item[1] A.Shojai and M.Golshani,Annales de la Fondation Louis de Broglie,
{\bf 22} (4) , 373 (1997).
\item[2] T.F. Jordan, Linear Operators for Quantum Mechanics (Wiley,
New York, 1969).
\item[3] T.F. Jordan, Am. J. Phys. {\bf 43}, 1089 (1975) .
\item[4] L.E. Ballentine, Quantum Mechanics (Prentice-Hall, 1990).
\item[5] H.Razmi and A.H.Abbassi,Tarbiat Modarres University (Dept. of Phys.)
Preprint(1998).
\end{itemize}

\end{document}